%% file: main.tex
\documentclass[conference]{IEEEtran}
\IEEEoverridecommandlockouts
\def\BibTeX{{\rm B\kern-.05em{\sc i\kern-.025em b}\kern-.08em
    T\kern-.1667em\lower.7ex\hbox{E}\kern-.125emX}}

\input{packages_macros}
\begin{document}


\pagestyle{plain} 








\author{
\IEEEauthorblockN{
  Dorin Pomian\IEEEauthorrefmark{1},
  Abhiram Bellur\IEEEauthorrefmark{1},
  Malinda Dilhara,
  Zarina Kurbatova
}
\IEEEauthorblockA{
  \textit{University of Colorado Boulder, USA},
  \textit{JetBrains Research, Serbia} \\
  dorin.pomian@colorado.edu,
  abhiram.bellur@colorado.edu,
  malinda.malwala@colorado.edu,
  zarina.kurbatova@jetbrains.com
}

\IEEEauthorblockN{
  Egor Bogomolov,
  Timofey Bryksin,
  Danny Dig
}
\IEEEauthorblockA{
  \textit{JetBrains Research, the Netherlands, Cyprus},
  \textit{University of Colorado Boulder, USA} \\
  egor.bogomolov@jetbrains.com,
  timofey.bryksin@jetbrains.com,
  danny.dig@colorado.edu
}
\IEEEauthorblockA{
  \IEEEauthorrefmark{1} These authors contributed equally to this work.
}
}



\title{Together We Go Further: LLMs and IDE Static Analysis for Extract Method Refactoring}

\maketitle
\begin{abstract}
Long methods that encapsulate multiple responsibilities within a single method are challenging to maintain. 
Choosing which statements 
to extract into new methods has been the target of many research tools. 
Despite steady improvements, these tools often fail to generate refactorings that align with developers' preferences and acceptance criteria.
Given that Large Language Models (LLMs) have been trained on large code corpora, 
if we harness their familiarity with the way developers form functions, we could suggest refactorings that developers are likely to accept. 

In this paper, we advance the science and practice of refactoring by synergistically combining the insights of LLMs with the power of IDEs to 
perform Extract Method (\EM).  
Our formative study on \extendedCorpusSize  \EM scenarios revealed that LLMs are very effective for   giving expert suggestions, yet they are unreliable: up to \llmReliabilityPercentageExtendedCorpus of the suggestions are \emph{hallucinations}. 
We designed a novel approach that removes hallucinations from the  candidates suggested by LLMs, then further enhances and ranks suggestions based on static analysis techniques from program slicing, and finally leverages the IDE to execute refactorings correctly. We implemented this approach in an IntelliJ IDEA plugin called \tool. 
We empirically evaluated \tool on a diverse corpus that replicates \extendedCorpusSize actual refactorings from open-source projects. We found  that \tool outperforms previous state of the art tools: \tool suggests the developer-performed refactoring in \ecBestToolRecall  of cases, improving over the recall rate of \JExtractBestRecall for previous best-in-class tools. 
Furthermore, we conducted firehouse surveys with \totalResponse industrial developers and suggested refactorings on their recent commits. \agreeratio of them agreed with the recommendations provided by \tool.
This shows the usefulness of our approach and ushers us into a new era when LLMs become effective AI assistants for refactoring.
\end{abstract}

\input{sections/01-introduction}
\input{sections/02-background}

\input{sections/research-methodology}
\input{sections/04-evaluation}

\input{sections/06-threats-to-validity}
\input{sections/05-relatedWork}
\input{sections/06-conclusion}
\IEEEtriggeratref{56}
\bibliographystyle{IEEEtran}
\bibliography{references} 

\end{document}

%% file: packages_macros.tex
\usepackage{cite}
\usepackage{amsmath,amssymb,amsfonts}
\usepackage{algorithmic}
\usepackage{graphicx}
\usepackage{textcomp}
\usepackage{xcolor}
\usepackage{amsmath,amsfonts}
\usepackage{algorithmic}
\usepackage{algorithm}
\usepackage{array}
\usepackage[caption=false,font=normalsize,labelfont=sf,textfont=sf]{subfig}
\usepackage{textcomp}
\usepackage{stfloats}
\usepackage{url}
\usepackage{verbatim}
\usepackage{graphicx}
\usepackage{xspace}
\usepackage{hyperref}
\usepackage{listings}
\usepackage{makecell}
\usepackage{booktabs}
\usepackage{algorithm}
\usepackage{caption}
\usepackage{threeparttable}
\usepackage{longtable}
\usepackage{tabularray}
\usepackage{cleveref}
\hypersetup{hidelinks}
\usepackage{color, colortbl}
\usepackage{balance}
\usepackage[inline]{enumitem}
\usepackage{calligra}
\usepackage{amsmath,amsthm}
\usepackage{tikz}
\usetikzlibrary{shapes.geometric, arrows}
\usetikzlibrary{positioning}
\usepackage{pifont}
\usepackage{tcolorbox}
\usepackage{makecell}
\usepackage{fancybox}

\makeatletter

\lstdefinestyle{mystyle}{
    backgroundcolor=\color{backcolour},   
    basicstyle=\footnotesize,
    commentstyle=\color{codegreen},
    keywordstyle=\color{magenta},
    numberstyle=\tiny\color{codegray},
    stringstyle=\color{codepurple},
    basicstyle=\ttfamily\footnotesize,
    breakatwhitespace=true,         
    breaklines=true,                 
    captionpos=t,                    
    keepspaces=true,                 
    numbers=left,                    
    numbersep=5pt,                  
    showspaces=false,                
    showstringspaces=false,
    escapeinside={*}{*},
    showtabs=false,                  
    tabsize=2,
    frame=none,
    postbreak=\mbox{\textcolor{red}{$\hookrightarrow$}\space},
}
\lstdefinestyle{tablecode}{
  basicstyle=\small,
  breaklines=true,
  backgroundcolor=\color{white},
  stringstyle=\color{codepurple},
  numbers=none,
  frame=none,
  basicstyle=\ttfamily\scriptsize,
  xleftmargin=0pt,
  framexleftmargin=0pt,
  framexrightmargin=0pt,
  framexbottommargin=0pt,
  framextopmargin=0pt,
  showspaces=false,  
  showtabs=false,   
  showstringspaces=false, 
  escapeinside={|}{|},
  mathescape=true,
  morekeywords={value},
  deletekeywords=int,
}

\definecolor{tearose}{rgb}{0.96, 0.76, 0.76}
\definecolor{ruddypink}{rgb}{0.88, 0.56, 0.59}
\definecolor{pastelred}{rgb}{1.0, 0.41, 0.38}
\definecolor{javapurple}{rgb}{0.5,0,0.35}
\definecolor{teagreen}{rgb}{0.82, 0.94, 0.75}
\definecolor{codegreen}{rgb}{0.1,0.9,0}
\definecolor{yellow-green}{rgb}{0.6, 0.8, 0.2}
\definecolor{codegray}{rgb}{0.5,0.5,0.5}
\definecolor{codepurple}{rgb}{0.58,0,0.82}
\definecolor{backcolour}{rgb}{0.95,0.95,0.92}
\definecolor{blueannoback}{RGB}{234,242,250}
\definecolor{myblue}{rgb}{0, 0, 0.75}
\definecolor{lightgray}{gray}{0.85}
\definecolor{Blue}{rgb}{0.1, 0.1, 0.2}

\newcommand{\smcode}[1]{\texttt{\fontsize{7.5}{8}\selectfont #1}}

\theoremstyle{definition}
\newtheorem{definition}{Definition}[section]

\newcommand{\tool}{\emph{EM-Assist}\xspace}
\newcommand{\toolit}{\textit{EM-Assist}\xspace}

\newcommand*\circled[1]{\tikz[baseline=(char.base)]{
            \node[shape=circle,draw,inner sep=0.8pt] (char) {#1};}}

\lstdefinestyle{javaStyle}{
    language=Java,
    basicstyle=\footnotesize\ttfamily, 
    keywordstyle=\color{blue},
    commentstyle=\color{green!40!black},
    stringstyle=\color{purple},
    numbers=left,
    numberstyle=\tiny\color{gray},
    stepnumber=1,
    numbersep=5pt,
    backgroundcolor=\color{white},
    showspaces=false,
    showstringspaces=false,
    frame=none,
    rulecolor=\color{black},
    tabsize=4,
    captionpos=b,
    breaklines=true,
    breakatwhitespace=false,
    title=\lstname,
    escapeinside={\%*}{*)},
    morekeywords={*,...}
}

\newcommand{\EM}{\textit{EM}\xspace} 
\newcommand{\EMs}{\textit{EM}s\xspace}

\newcommand{\CommunityCorporaA}{Community~Corpus\xspace} 
 
\newcommand{\ExtendedCorpus}{Extended Corpus\xspace} 
\newcommand{\extendedCorpusProjects}{12\xspace}

\newcommand{\rqOnePrevDataset}{122\xspace}

\newcommand{\Comparisontools}{6\xspace}

\newcommand{\responserate}{80\%\xspace}
\newcommand{\agreeratio}{81.3\%\xspace}
\newcommand{\totalSent}{20\xspace}
\newcommand{\totalResponse}{16\xspace}

\newcommand{\SA}{2\xspace}
\newcommand{\AG}{6\xspace}
\newcommand{\SWA}{5\xspace}
\newcommand{\SWD}{0\xspace}
\newcommand{\D}{1\xspace}
\newcommand{\SD}{2\xspace}

\newcommand{\extendedCorpusSize}{{1752}\xspace}

\newcommand{\llmReliabilityPercentageExtendedCorpus}{76.3\%\xspace}
\newcommand{\llmFinalUsefullPercentageExtendedCorpus}{23.7\%\xspace}
\newcommand{\mean}{mean\xspace}
\newcommand{\llmInvalidSuggestionsPercentageExtendedCorpus}{57.4\%\xspace}
\newcommand{\llmNotUsefulSuggestionsPercentageExtendedCorpus}{18.9\%\xspace}

\newcommand{\bestJetGPTRecall}{60.6\%\xspace}

\newcommand{\bestMLRecall}{54.2\%\xspace}

\newcommand{\bestStaticAnalysisRecall}{52.2\%\xspace}

\newcommand{\llmNumberOfSuggestionsProducedOnExtendedCorpus}{22741\xspace}

\newcommand{\promptExampleSize}{2\xspace}
\newcommand{\codeFragmentUpperThreshold}{88\%\xspace}
\newcommand{\averageUsefulSuggestionsPerFunctionCommunityCorpora}{10\xspace}
\newcommand{\averageSuggestionsPerFunctionCommunityCorpora}{27\xspace}

\newcommand{\errOrHallSuggestionsExtendedCorpus}{17356\xspace}
\newcommand{\applicableSuggestionsExtendedCorpus}{5385\xspace}

\newcommand{\bestJetGPTOverBestJExtractRecall}{35\%\xspace}

\newcommand{\JExtractBestRecall}{39.4\%\xspace}

\newcommand{\meFixPointIteration}{9\xspace}
\newcommand{\meSuggestionsSize}{9\xspace}
\newcommand{\meApplicableSuggestionsSize}{3\xspace}
\newcommand{\meInvalidSuggestionsSize}{3\xspace}
\newcommand{\meNotUsefulSuggestionsSize}{3\xspace}
\newcommand{\meHallucinationSuggestionsSize}{6\xspace}
\newcommand{\meOracle}{1\xspace}

\newcommand{\ecBestToolRecall}{53.4\%\xspace}

\newcommand{\emToolRecallOne}{57.6\%\xspace} 
\newcommand{\emToolRecallTwo}{58.4\%\xspace} 
\newcommand{\emToolRecallThree}{63\%\xspace} 

\newcommand{\resultbox}[1]{
\begin{center}
\begin{tcolorbox}[colback=white,colframe=gray!60!black,left=0pt,right=0pt,top=0pt,bottom=0pt]
    \begin{minipage}{.99\textwidth}
    #1
    \end{minipage}
\end{tcolorbox}
\end{center}
}

%% file: sections/01-introduction.tex
\section{Introduction\label{sec:introduction}}

Excessively long methods that encapsulate multiple responsibilities within a single method are challenging to comprehend, debug, reuse, and maintain~\cite{Tsantalis2011,banker1993software,fowler1997refactoring}. 
To mitigate these issues, software developers use Extract Method (\EM)  refactoring -- a hallmark refactoring~\cite{fowler1997refactoring} supported by all modern IDEs.
This refactoring involves moving a block of statements from a host method to a brand new method, passing the used variables as parameters to the new method, and adding a call to the new method from the host method. This refactoring is reported~\cite{negara2013,murphy2012,refactoringminer2} as being among the top-5 most frequently performed in practice, both for manual and automated refactoring.


The process of performing an \EM consists of two phases: (i) choosing the statements to extract and (ii) applying the mechanics to perform this refactoring. 
While the application part has been a staple feature of all modern IDEs, they leave it up to developers to choose the statements to extract.
Researchers developed several tools that suggest which statements to extract based on static analysis~\cite{Maruyama2001,Tsantalis2011,Charalampidou2017,Yang2009APSEC,Tiwari2022ISEC}, or machine learning (ML) models~\cite{gems2017ISSRE,REMS2023ICPC}. 
These tools use software quality metrics (e.g., coupling, cohesion), which they optimize  based on heuristics or by training ML classifiers.


While these tools adhere to software engineering principles like the Single Responsibility Principle~\cite{cleanArchitecture}, they fail to align with 
real-world \EM instances. 
We posit that the decision to refactor is both \emph{a science and an art}. Developers use both their knowledge of software engineering principles \emph{and} rely on their own experience and subjective interpretation of the code context and what makes a good method.  
This might explain why developers are reluctant to use automated refactorings~\cite{WhyWeRefactorFSE2016} and the large gap between high metrics scores and their low acceptance by developers \cite{fakhoury2019improving,scalabrino2019automatically,software2,mlcodechanges,pyevolve}.

Recently, Large Language Models (LLMs)~\cite{feng2020codebert, gpt-4, google_bard} are emerging as powerful companions for several software engineering tasks~\cite{Matteo2022TSE,feng2020codebert,CITE_AleksandraASE23,Feng2024ICSE,dilharaFSE2024LLM4TBE}.
Given LLMs' training on extensive code repositories that contain billions of methods written by actual developers,
we hypothesize that they are more likely to imitate human behavior by mimicking how developers form functions, thus are likely to suggest refactorings that developers would accept.
{Our formative study on \extendedCorpusSize \EM scenarios from \extendedCorpusProjects open-source projects  revealed that LLMs are very effective in giving expert suggestions. Moreover, they are very prolific, producing on average \averageSuggestionsPerFunctionCommunityCorpora suggestions per method. 
However, we also discovered that LLMs are unreliable: {\llmReliabilityPercentageExtendedCorpus} of their suggestions are \emph{hallucinations}, i.e., they seem plausible at first, but are actually deeply flawed. 
We found that \llmInvalidSuggestionsPercentageExtendedCorpus of the suggested refactorings contain code fragments that are invalid to extract (e.g., would produce compile errors), and \llmNotUsefulSuggestionsPercentageExtendedCorpus are not useful (e.g., suggest to extract the whole host method).}


To advance the state of the art and practice for refactoring, we bridge several important gaps. First, we bridge the gap between the suggestion of refactorings and developers' actual practices by grouping statements into methods that resemble human-written code. Second, we bridge the gap between suggesting and applying refactorings by supporting the whole end-to-end process in a way that provides maximum automation while taking into account human input. 

We have designed, implemented, and evaluated \tool, an IntelliJ IDEA plugin, that supports Java and Kotlin \EM refactorings. \tool synergistically combines (i) the creative capabilities of LLMs, (ii) static analysis techniques to filter and enhance LLM-provided suggestions, and (iii) the full power of a state-of-the-practice commercial IDE to apply refactorings safely. \tool first repeatedly prompts the LLM in a few-shot learning style to generate a diverse range of refactoring suggestions. Subsequently, it eliminates two kinds of hallucinations: (i) it employs static analysis techniques from the IDE to eliminate invalid suggestions (i.e., illegal groupings of code statements), and (ii) it eliminates suggestions that are not useful (e.g., that include the whole body of the host method, or one single line). Then it further enhances the remaining valid suggestions based on program slicing techniques. 
Following this step, it ranks and prioritizes high-quality suggestions so that it would not overwhelm the developer with too many suggestions. 
It then presents the ranked suggestions to the developers. Lastly, it encapsulates the user-chosen candidate into a refactoring command and uses the IDE to correctly execute the refactoring.

We designed a comprehensive empirical evaluation to determine the benefits of our novel approach, the pros and cons of using LLMs, and a \emph{sensitivity analysis}. 
To determine the effectiveness of \tool, we use two complementary methods. First, we use a publicly available corpus~\cite{gems2017ISSRE} of \rqOnePrevDataset \EMs that other researchers used in the past. 
The results show that our \tool outperforms state-of-the-art static-analysis tools such as JDeodorant~\cite{Tsantalis2011}, JExtract~\cite{jextractsilva2015}, SEMI~\cite{Charalampidou2017}, LiveRef~~\cite{fernandes2022live, liferef2022ASE}, and also outperforms ML-based techniques like REMS~\cite{REMS2023ICPC} and GEMS~\cite{gems2017ISSRE}. \tool correctly suggests the ground truth \EM refactoring among the top-5 candidates \bestJetGPTRecall of times, compared to \bestMLRecall reported by
existing ML models, and \bestStaticAnalysisRecall reported by existing static analysis tools.
Second, we significantly expanded the previous community corpus of \rqOnePrevDataset \EMs 
with a diverse corpus of \extendedCorpusSize \emph{actual} \EM instances
from open-source projects. \tool correctly suggests the actual refactoring performed by the developers in \ecBestToolRecall
of cases, an improvement over the \JExtractBestRecall rate of the previous best-in-class static analysis tool. 

Moreover, to assess whether refactoring suggestions generated by \tool are useful, we employed firehouse surveys~\cite{firehouse2015TSE} with \totalResponse industrial developers from a reputable software company. 
We presented them with refactoring suggestions for lengthy methods they previously committed into their software repository.
\agreeratio of respondents agreed that the suggestions were beneficial and suitable for application in their codebase. They wish to see \tool in production in IntelliJ IDEA and use it in their daily coding. 
This demonstrates that \tool generates suggestions that are highly likely to be embraced by developers and ushers us into a new era of refactoring when developers are accepting AI assistants.


In summary, this paper makes the following contributions:
\begin{enumerate}[label=\bfseries (\arabic*),wide, labelindent=0pt,labelwidth=0.1mm]
\item  
We present the first approach that automates the full refactoring lifecycle by synergistically combining LLMs and IDEs. 
Thus, we effectively bridge the gap between refactoring recommendation and application, and we shrink the gap between refactoring suggestions and developer practices.

\item We discovered best practices for prompting and tuning LLMs to produce effective refactoring suggestions and we reveal LLMs' strengths and weaknesses for refactoring.

\item We designed, implemented, and evaluated \tool, a plugin for IntelliJ IDEA, that integrates all these ideas.

\item Our comprehensive empirical evaluation on a corpus used by others, as well as significantly extended corpus that replicates \extendedCorpusSize real-world refactorings shows that \tool outperforms static analysis-based techniques and ML approaches that suggest \EM refactoring. Moreover, our survey with \totalResponse industrial developers provides insights about the reasons why developers accept or reject \tool's suggestions. 
\end{enumerate}

To aid reproducibility, we make \tool, a  demo, and all our experimental data available anonymously~\cite{replicationPackage}.

%% file: sections/02-background.tex
\section{Motivating Example}\label{sec:background}



\begin{figure}
    \includegraphics[width=0.5\textwidth,keepaspectratio]{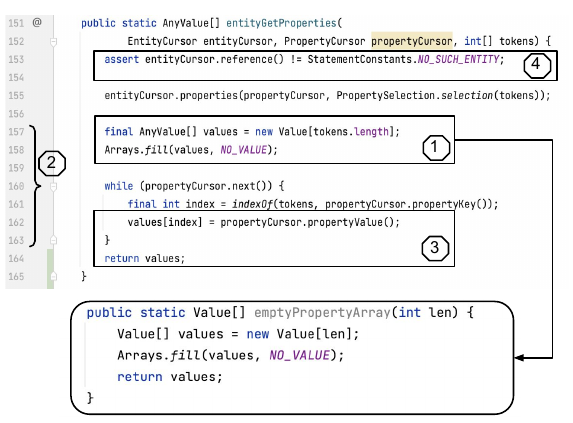}
    \caption{The numbered code snippets represent (1) an extract function refactoring in the project Neo4j, commit a05a8c5, (2) a suggestion made by static analysis tool, 
    (3) an invalid suggestion from LLM, (4) a not useful suggestion from LLM.
    }
    \label{fig:motivation}
    \vspace{-2mm}
\end{figure}

We illustrate the challenges of suggesting code fragments to extract into new methods in a way that aligns with developer preferences, as well as the process of unleashing the full potential of LLMs for \EM refactoring. 
\Cref{fig:motivation} shows an \EM  refactoring performed by Neo4J developers. 
They extracted statements in lines 157 and 158 into a new method named \smcode{emptyPropertyArray} (\circled{1} in \Cref{fig:motivation}). 
This refactoring decision is intuitively sound as these two statements collectively fulfill a single responsibility, namely, the creation of an empty array.

When replicating this real-world scenario, existing techniques~\cite{Maruyama2001,Tsantalis2011,Charalampidou2017} for suggesting \EM would recommend extracting lines 157--163 (\circled{2} in \Cref{fig:motivation}). 
This recommendation arises from the fact that the variable \smcode{values} is used later in the code following the two aforementioned statements. 
However, this selection of statements does not align with the developers' actually performed refactoring (\circled{1} in \Cref{fig:motivation}) despite the fact that the extraction of lines 157-163 adheres to the static analysis principles (e.g., extracting a program slice) and software-quality metrics (e.g., improving the cohesion of statements in a method) governing existing tools. This finding is in line with research that shows that code-quality metrics do not necessarily capture code quality improvements as perceived by the developers~\cite{fakhoury2019improving,scalabrino2019automatically}.

Next, we used a LLM to suggest \EM refactorings for this method, specifically opting for GPT-3.5~\cite{gpt-3} because it is cost-effective.
We provided the LLM with a prompt that included the code snippet in the host method \smcode{entityGetProperties}, followed by a description of the \EM refactoring process and an instruction to generate refactoring suggestions for the method.
As LLMs produce results that are non-deterministic, we repeated the same prompting and collected suggestions until we reached a fix-point, i.e., the LLM no longer produced new suggestions. We reached the fix-point in \meFixPointIteration iterations. 

The LLM came up with 
\meSuggestionsSize distinct suggestions for which statements to extract, among these including the one that the Neo4J developer performed. 
In addition to 
\meApplicableSuggestionsSize applicable suggestions, 
the LLM also proposed to extract lines 162--164 (\circled{3} in \Cref{fig:motivation}). If we did so, the code would not even compile because the suggestion starts inside the \smcode{while} loop and goes past the closing bracket. We call these suggestions \emph{invalid}. The LLM produced \meInvalidSuggestionsSize invalid suggestions. While the LLM might have seen those code statements being associated together in codebases, because the LLM does not fully understand the semantics of \EM and syntactical rules of well-formed code, those associations render the extracted method not compilable.

Moreover, the LLM suggested extracting the code fragment in line 153 (\circled{4} in \Cref{fig:motivation}), which is a one-liner. While it is possible to extract this line correctly, we consider it \emph{not-useful} as it is improbable for developers to undertake this suggestion. Similarly, the LLM suggested to extract the whole method body into a new method, which does not provide any value for the developers. We call such suggestions \emph{not-useful}. In this experiment, the LLM produced \meNotUsefulSuggestionsSize not-useful suggestions. The invalid suggestions, together with the not-useful ones, represent \emph{hallucinations} of the LLM.

This experiment showcases the strengths of LLMs: being prolific in generating \meSuggestionsSize suggestions, among those is \meOracle suggestion that aligns with the way how the developers actually performed the refactoring. 
However, it also showcases the LLM's weakness: it produced \meHallucinationSuggestionsSize hallucinations, of which \meInvalidSuggestionsSize are invalid suggestions and \meNotUsefulSuggestionsSize are not useful. 
This shows that it is imperative to exercise caution and not exclusively rely on LLM-generated suggestions when conducting refactorings assisted by LLMs. Moreover, it shows that while LLMs have a huge potential to come up with refactorings that match human's acceptance criteria, developers that embark on such a journey need to work hard to tap into such potential: they need to tame LLMs non-determinism by repeated prompting (\meFixPointIteration in our example), collect those \meSuggestionsSize suggestions, and sift through and discard \meHallucinationSuggestionsSize hallucinations. Moreover, as we show in~\Cref{sec:rq2}, developers might need to repeat the whole suite of experiments for different LLM ``temperature'' settings: these control the level of variability for the solutions produced by the LLM. Furthermore, developers might want to use various LLMs (in~\Cref{sec:rq1} we tried three LLMs), altogether resulting in a combinatorial explosion of experiments.  

This is where our tool, \tool, liberates the developers so they can focus on the creative part: using their expertise to examine a small number of high-quality suggestions. 





%% file: sections/research-methodology.tex
\section{Technique}
\begin{figure}
    \includegraphics[width=0.45\textwidth,keepaspectratio]{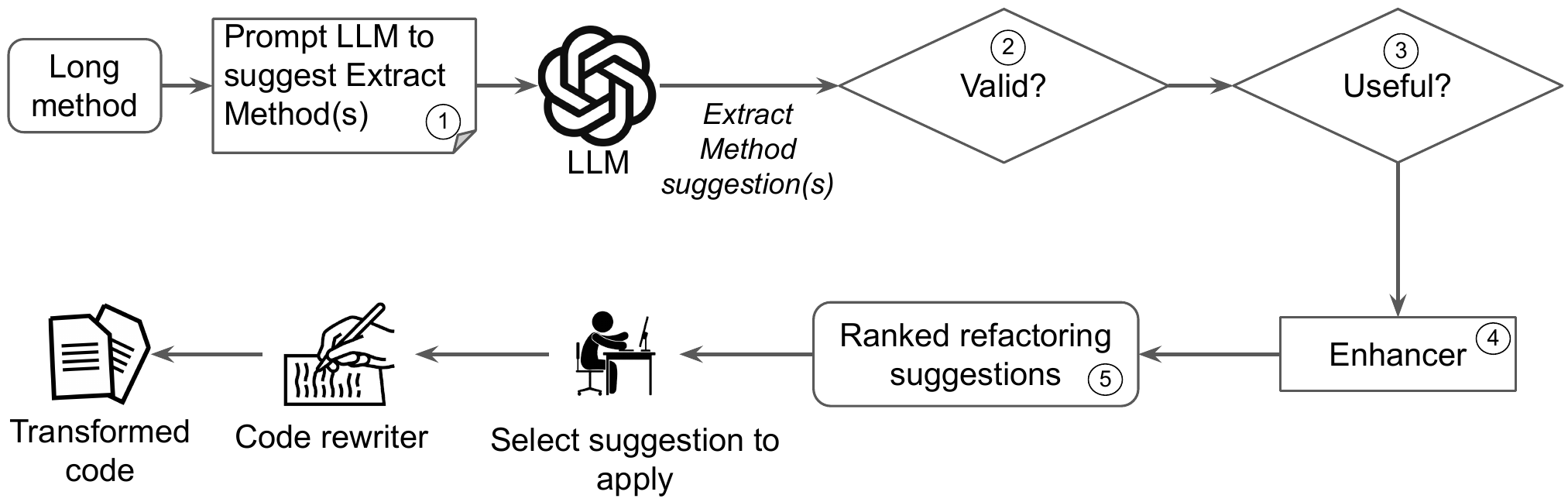}
    \caption{The workflow of generating refactoring suggestions and then applying them with \tool.}
    \label{fig:workflow}
    \vspace{-2mm}
\end{figure}

In this section, we present the workflow that our novel approach and tool, \toolit, uses to automatically suggest and perform \EM refactoring in Java and Kotlin codebases.  

\Cref{fig:workflow} shows the architecture and the steps performed by \toolit.
First, \tool invokes the LLM (\circled{1} in \Cref{fig:workflow}) by prompting it to generate \EM suggestions for the selected method (see \Cref{sec:get_suggestions}).
Next, our algorithm removes hallucinations and retains \emph{applicable} refactoring options that are free of 
errors (\circled{2} in \Cref{fig:workflow}, detailed in \Cref{sec:get_valid}) and are useful (\circled{3} in \Cref{fig:workflow}, detailed in \Cref{sec:get_useful}). 
These steps play a key role in retaining suggestions that align with the way developers extract code, while simultaneously mitigating the risk of introducing bugs or non-compilable code. 

Next, our algorithm further \emph{enhances} (\circled{4} in \Cref{fig:workflow}, detailed in \Cref{sec:enhancing}) the quality of the suggestions by expanding or shrinking the code fragment in the suggestion based on program slicing rules. These rules ensure that refactorings better align with the developer's intentions.
Then, the tool ranks the refactoring suggestions (\circled{5} in \Cref{fig:workflow}, detailed in \Cref{sec:ranking}) to prevent overwhelming the developer, presenting only the top-n ranked options for the developer to choose from.
Tool's GUI displays the suggestions and shows a preview of the signature of the extracted method, along with the code fragment that would be extracted. Based on the user's decision, \tool encapsulates the suggestion in a refactoring command and invokes IntelliJ IDEA to perform the refactoring correctly. 


\begin{definition}{(Long method\label{def:long_method})}
     is chosen by a developer for extract method refactoring, denoted as $l_i(s_i,e_i)$. The method starts at line $s_i$, where its signature is defined, and ends at line $e_i$ containing the closing curly bracket that signifies the end of the method declaration. The method's length, $L_i$, is $e_i - s_i$.
\end{definition}

\begin{definition}{(Extract Method\label{def:extract_method})}
denoted as \(e_j(n_j^e, (s_j^e, e_j^e)\), where \(n_j^e\) represents the method name, \((s_j^e, e_j^e)\) signifies a code fragment within a long method. 
It is identified by the starting line number, defined as \(s_j^e > s_i\), and the ending line number, defined as \(e_j^e < e_i\). These line numbers are calculated with respect to the original source code file in which the host method $l_i(s_i,e_i)$ is located.
\end{definition}

\begin{definition}{(Extract Method Suggestions ($\mathcal{S}$)\label{def:extract_method_suggestions})}
is a set of extract method suggestions generated by an LLM for a long method $l_m$, formally,
$\mathcal{S} = \{ e_i(n_i, (s^e_i, e^e_i)), e_j(n_j, (s^j_s, e^j_e)), \ldots \}$


 \end{definition}
 
\begin{definition}{(Invalid Extract Methods ($\mathcal{I})$\label{def:invalid_extract_method})} 
 is a subset of the $\mathcal{S}$ suggestions, such that it does not satisfy the validity conditions ($C(e_i)$), ensuring that the resulting code remains compilable. Formally, $\mathcal{I}_i = \{e_i \in \mathcal{S} \mid \neg C(e_i)\}$.
The remaining set, $\mathcal{V}$, comprises suggestions that satisfy the validity conditions, making them valid suggestions for use. Formally, $\mathcal{V}_i = \{e_i \in \mathcal{S} \mid C(e_i)\}$.
 
 \end{definition}


\begin{definition}{(Not Useful Extract Methods $(\mathcal{NU})$\label{def:notuseful_extract_method})}
is a subset of $\mathcal{V}_i$ comprising elements that fail to meet the criteria for usefulness, $U(e_i)$. This subset ensures that the suggestions are neither too large, encompassing almost the entire method body, nor too small, essentially one-liners.
Formally, $\mathcal{NU}_i = {e_i \in \mathcal{V} \mid \neg U(e_i)}$.
Conversely, useful suggestions $\mathcal{U}_i$, consist of elements satisfying the usefulness criteria $U(e_i)$, suitable for method extraction, defined as $\mathcal{U}_i = {e_i \in \mathcal{V} \mid U(e_i)}$.
\end{definition}

\subsection{Generating \EM Suggestions\label{sec:get_suggestions}}
We utilize LLMs to generate \EM recommendations for a given input long method. In this section, we delve into the preparation of LLM for \EM refactoring suggestions, as well as the parameters of the LLM that require tuning.


\subsubsection{\textbf{Prompt engineering}}
LLMs are versatile models capable of various applications, but they require specific preparation when used for a particular task~\cite{Feng2024ICSE}. 
To facilitate this, we employ \emph{in-context learning}~\cite{NEURIPS2022_9d560961}, where we provide all the instructions needed to solve the task right in the model's input, including task definition and relevant context information. 
To further enrich the prompt, we employ few-shot learning~\cite{gpt-3,radford2019language} by incorporating a set of \promptExampleSize examples into the prompt, following best practices as discussed by Gao et al.~\cite{Gao:2023ASEIn-context}.
The prompt includes:
\begin{enumerate*}[label=(\roman*)]
\item A succinct overview of \EM refactoring,
\item The definition of a long method,
\item A JavaDoc string when available,
\item  \promptExampleSize examples of refactoring for one long and one short method, and
\item Precise instructions for the desired output format.
\end{enumerate*}
{An example prompt is accessible on our companion website~\cite{prompt}}.

\subsubsection{\textbf{Generating an extensive array of suggestions}\label{sec:gen_extensive}}
The output of an LLM depends on two factors:
\begin{enumerate*}[label=(\roman*)]
\item the internal variable ``Temperature'' ($T$), and
\item the number of iterations ($I$)~\cite{gpt-3, liu2019roberta, radford2019language}, which indicates how many times the same prompts are inputted.
\end{enumerate*}
Temperature serves as a regulator for the model's output randomness. 
Higher values, such as 0.9 or 1.2, produce more diverse and unpredictable outputs, whereas values closer to 0 produce focused and less diverse results. 
We determine the optimal value empirically (see \Cref{sec:rq2}).

Even when identical prompts are presented multiple times, the output can exhibit variations due to: 
\begin{enumerate*}[label=(\roman*)]
\item LLMs employ a combination of learned patterns and random sampling in generating responses, introducing slight deviations in their answers each time due to inherent randomness; 
\item LLMs possess the capability to explore various potential solutions and refine their responses iteratively through interactions and feedback~\cite{gpt-3, radford2019language}, continually enhancing their outputs in subsequent iterations.
\end{enumerate*}
Therefore, the quality and quantity of predictions made by an LLM are influenced by the frequency of prompting the same prompt.

In \Cref{sec:rq2}, we conduct a \emph{sensitivity analysis} to determine how the performance of \tool varies with each combination of temperature and iterations. This analysis enables \tool to use the optimal settings for generating refactoring suggestions.


\subsection{Removing Invalid \EM Suggestions\label{sec:get_valid}}
Not all suggestions generated by LLMs can be directly applied to codebases, as some may lead to non-compilable code. Therefore, \tool analyzes each suggestion to ensure it can be extracted without breaking the code's scope. 
For example, suggestion \circled{3} in our motivating example (\Cref{fig:motivation}) fails the scope analysis, and \tool expands the code fragment of the suggestion so that both the start and end lines have the same scoping level: the new suggestion's code fragment starts from line 160, while the end line stays the same. This change, however, leads to a compilation error.

To remove suggestions that might lead to non-compilable code, \tool uses the rules below: 
\begin{enumerate}[wide, leftmargin=*, labelwidth=!, labelindent=0pt, label=(\roman*)]
    
    \item \textit{Variable Usage:} This criterion ensures that all necessary variables and objects are accessible within the same scope or are provided as parameters, maintaining the code's self-containment.
    
    \item \textit{Return Values:} If the code being extracted produces a result or modifies the state of objects, \tool validates that the return value and side effects are appropriately handled. For example, it checks if the return value is used subsequently in the host method or if any thrown exceptions are caught.
    
    \item \textit{Number of Return Values:} Java methods can return only one value, and hence, \tool validates whether the code fragment to be extracted might produce more than one result, in which case it discards the suggestion. 
    
    
    \item \textit{Control Flow:} \tool reasons about the control flow within the extraction suggestion to ensure that it can be extracted as a separate method without causing logical errors. For example, it checks for any early returns or breaks that might affect the control flow and would render the extracted method to have a different behavior than the original code.
\end{enumerate}
\noindent
\tool checks all these conditions through the static analysis infrastructure in the IntelliJ Platform\footnote{The IntelliJ Platform: \url{https://www.jetbrains.com/opensource/idea/}.} (the open-source platform IntelliJ IDEA and other JetBrains IDEs are built upon) that verifies refactoring preconditions. As LLMs are not aware of the semantics of \EM refactoring, but only about the code tokens that are associated together in its training data, this step is crucial for discarding a large number of invalid suggestions (\llmInvalidSuggestionsPercentageExtendedCorpus of all suggestions).


\subsection{Removing Refactoring Suggestions That Are Not Useful}\label{sec:get_useful}

Once we retain only valid refactoring suggestions, the next step filters out those suggestions that are not useful 
and retains the ones that are applicable and likely to be performed by a developer.
To identify suggestions that are not useful, we employ two rules: 
\begin{enumerate*}[label=(\roman*)]
\item Exclude \EM suggestions that encompass \codeFragmentUpperThreshold or more of the statements present in the long method, and 
\item Following practices adopted by other researchers \cite{Tsantalis2011, fernandes2022live, haas2015deriving,yang2009identifying}, we exclude \EM suggestions that are one-liners. 
\end{enumerate*}

Extracting an entire method or a one-liner into another method is, in general, not useful for the purpose of improving existing code. These might be useful if the developers plan to add new code that implements new features into the host method or into the extracted method. Notice that we designed \tool for \emph{code renovation} and not for feature expansion.

\subsection{Enhancing Refactoring Suggestions\label{sec:enhancing}}
After identifying useful suggestions (\Cref{def:notuseful_extract_method}) devoid of hallucinations, the next step enhances these suggestions to further improve the quality of valid suggestions.

\begin{enumerate}[wide, leftmargin=*, labelwidth=!, labelindent=0pt, label=(\roman*)]

\item \textit{Program slicing}:
Taking inspiration from tools that rely solely on static analysis, such as those utilizing program slicing~\cite{Maruyama2001,Tsantalis2011,Charalampidou2017,abadi2009fine,meananeatra2018refactoring}, we leverage program slicing to augment refactoring suggestions provided by LLMs. We use rules that better align with developer preferences to avoid creating small methods with several arguments.


Let $EF_s$ be the initial statement of the code fragment to be extracted into a new method, with $EF_{s+i}$ indicating statements below $EF_s$ and $EF_{s-i}$ denoting statements above $EF_{s}$ within the long method.
Let $EF_{s-1}$ represent the variable declaration statement for variable $v$, and this variable is used inside the code fragment of the suggested method extraction, identified through live variable analysis based on a def-use chain.
In such cases, we increase the scope of the code to be extracted by one statement, including $EF_{s-1}$ in the extracted method, and we adjust the starting line of the suggestion to match the starting line of $EF_{s-1}$.
This avoids having to pass $v$ as a parameter. 




\item \textit{Control statements}:
If the code fragment in the extracted method starts with a control statement (i.e., $EF_{s}$ is the check block of an \smcode{if} statement), we shrink the code fragment to start at $EF_{s+1}$ and contain only the block of the \smcode{if}. This leaves the check condition intact in the host method so that the developer can see under which conditions the would-be extracted method is called, thus making the code easier to read. 
\end{enumerate}

We evaluate the effectiveness of these heuristics in~\Cref{sec:rq2}.
We plan to experiment with other heuristics based on the feedback we receive on JetBrains Marketplace~\cite{tool:Marketplace}.

\subsection{Ranking Refactoring Suggestions\label{sec:ranking}}

The number of the remaining applicable suggestions per host method can be large (on average 
\averageUsefulSuggestionsPerFunctionCommunityCorpora
 per method) and overwhelm the developer. 
To prioritize high-quality suggestions, we implemented a ranking mechanism that gives precedence to recommendations consistently identified by the LLM. 
This approach elevates frequently highlighted suggestions, employing the LLM's extensive knowledge of coding best practices.
Next we present our ranking function.

\noindent\scalebox{0.8}{
\parbox{\linewidth}{
\begin{align}
    \mathcal{T}_1(e_i) & = \mathcal{H}(e_i) = \sum_{i=1}^{N} \mathcal{F}(line_i) \\
    \mathcal{T}_2(e_i) & = \mathcal{P}(e_i) \\ 
    \mathcal{T}_3(e_i) & = \mathcal{H}(e_i) \cdot \mathcal{P}(e_i)
\end{align}
}}

\noindent\scalebox{0.8}{
\parbox{\linewidth}{
Where:
\begin{align*}
    N : & \text{Total number of lines in the \EM suggestion.} \\
    \mathcal{F}(line_i) : & \text{Number of times the $i^{th}$ line appears in all the suggestions}.\\
    \mathcal{H}(e_i) :& \text{Heat of the \EM suggestion } e_i. \\
    \mathcal{P}(e_i) :& \text{Popularity of the \EM suggestion } e_i. \\
\end{align*}
}}


Our approach begins with $\mathcal{T}_1(e_i)$, where we assign scores to each suggestion according to a "heat map" of the host method. 
Intuitively, this ranking measures the LLM's confidence that a  certain line of code in the host method belongs to another method. 
To compute the heat map, we record how many times each line appears in all applicable suggestions. 
If a line is absent from all suggestions, we assign it a score of zero. 
We repeat this procedure for every line within the host method. 
Then, we aggregate individual line scores to determine each suggestion's overall heat score, comprising multiple code lines, and then rank these suggestions based on their heat scores.

$\mathcal{T}_2(e_i)$ ranks suggestions by considering their popularity.
As discussed in \Cref{sec:gen_extensive}, we re-prompt the LLM several times with the same prompt.
While we remove duplicated suggestions, we keep track of how many times a certain suggestion is produced by the LLM during re-prompting. 
Given that LLMs have been trained on extensive codebases enabling them to mimic how real developers construct methods, we give precedence to suggestions that appear repeatedly. 

Finally, $\mathcal{T}_3(e_i)$ combines both the heat and popularity of suggestions using a weighted average, aiming to strike a balance between the importance of these two factors.
We evaluate the effectiveness of the rankings in \Cref{sec:rq2}.

%% file: sections/04-evaluation.tex
\section{Evaluation}\label{sec:evaluation}
We empirically evaluate \toolit by answering the following research questions:
\begin{enumerate}[label=\bfseries RQ\arabic*.,wide, labelwidth=!, labelindent=0pt]
\item \textbf{How effective are LLMs at generating refactoring suggestions?}
Understanding the capability of LLMs to produce 
\EM suggestions is essential for our tool's effectiveness. Therefore, we conducted a quantitative analysis involving three of the latest LLMs available.

\item \textbf{How do refactoring suggestions change with different LLM parameters?} 
This is important for integrating LLMs into tools and for researchers using LLMs for refactoring. 
We conduct a \emph{sensitivity analysis} to assess the quality of these suggestions, focusing on the recall rate for the top-5 suggestions (Recall@5) across various LLM parameters, and examine the effects of our enhancements and ranking methods.
\item \textbf{How effective is \tool in providing refactoring recommendations over existing approaches?} 
To quantitatively evaluate this, we conducted a baseline comparison with \Comparisontools other tools that employ static analysis or machine learning techniques to suggest \EM refactorings. 
\item \textbf{How useful are the provided recommendations to developers?}
\tool is a code renovation tool that developers use interactively. Thus, it is important to evaluate whether \tool makes suggestions that developers accept. 
We employ firehouse surveys with professional developers from our collaborating enterprises, focusing on newly created long methods they committed into code repositories. 
\end{enumerate}

\subsection{Datasets}\label{sec:datasets} 
To answer our research questions, we use two datasets: the ``Community Corpora'' previously used by other researchers working in this field, and the ``Extended Corpus'' which we released to address the limitations of previous corpora and to increase its size. 

\begin{enumerate}[label=(\roman*),wide, labelwidth=!, labelindent=0pt]
\item \textbf{\textit{\CommunityCorporaA:}}
consists of \rqOnePrevDataset Java methods and their corresponding \EM refactorings using five open-source repositories: MyWebMart, SelfPlanner, WikiDev, JHotDraw, and JUnit. 
This dataset previously served as the foundation for evaluating various state-of-the-art \EM refactoring automation tools, including \textit{JExtract}~\cite{jextractsilva2015}, \textit{JDeodorant}~\cite{Tsantalis2011}, \textit{SEMI}~\cite{Charalampidou2017}, \textit{GEMS}~\cite{gems2017ISSRE}, and \textit{REMS}~\cite{REMS2023ICPC}. 
The \CommunityCorporaA, although extensively used by researchers, can be seen as subjective. It combines (i) theoretical refactoring scenarios assessed by third-party (usually students) who are not familiar with the open-source projects, and (ii) synthetic refactorings devised by corpus designers, based on inlining existing methods followed by extracting these same methods from the expanded host methods. 
While these refactorings are realistic, they are not actual refactorings that developers performed. Thus, we created another corpora that addresses these limitations.


\item \textbf{\textit{\ExtendedCorpus:}}
To further strengthen our evaluation's robustness with a broad oracle of actual refactorings by developers and ensure more generalizable results, we constructed \ExtendedCorpus. To create it, we employed RefactoringMiner~\cite{refactoringminer2}, the state-of-the-art tool for mining refactorings from commits, with a reported precision of 99.8\% and recall of 95.8\% for detecting \EM. 
We ran RefactoringMiner on \extendedCorpusProjects open-source repositories, including notable projects like CoreNLP, Guava, and Gson, which span a wide range of domains such as machine learning, database management, data handling, and web technologies.
We filtered out refactoring commits that include one-liners and extracted methods whose bodies overlapped significantly with the host method. 
Additionally, we filtered out non-automatable~\cite{Cossette2012FSE} refactorings, such as those entailing feature additions. 
For this purpose, we leveraged the state of the practice IDE for \EM, IntelliJ IDEA, using start and end line numbers of the extracted code fragment provided by RefactoringMiner to determine whether IntelliJ IDEA could automatically extract the code fragment identified. 

Consequently, we retained \extendedCorpusSize \EMs from these repositories. We demonstrate that this corpus is representative in terms of the length of the host method, cyclomatic complexity~\cite{McCabeComplexity} of the host method, and Halstead metric difficulty~\cite{Halstead} of the host method, as shown in \Cref{fig:extended_corpus_metrics}.

\begin{figure}[ht]
  \centering
  \begin{minipage}{.33\columnwidth}
    \includegraphics[width=\linewidth]{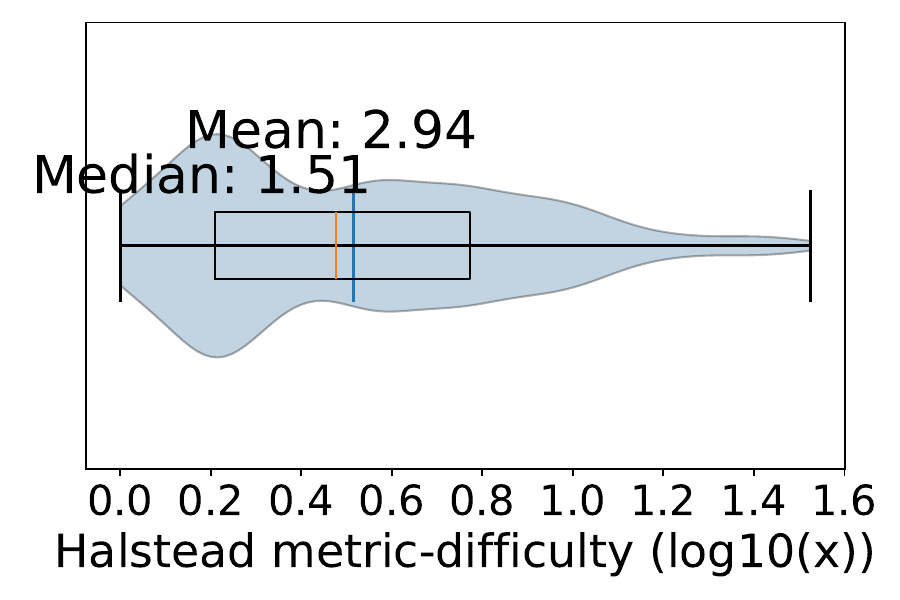}
  \end{minipage}
   \hfill 
  \begin{minipage}{.32\columnwidth}
    \includegraphics[width=\linewidth]{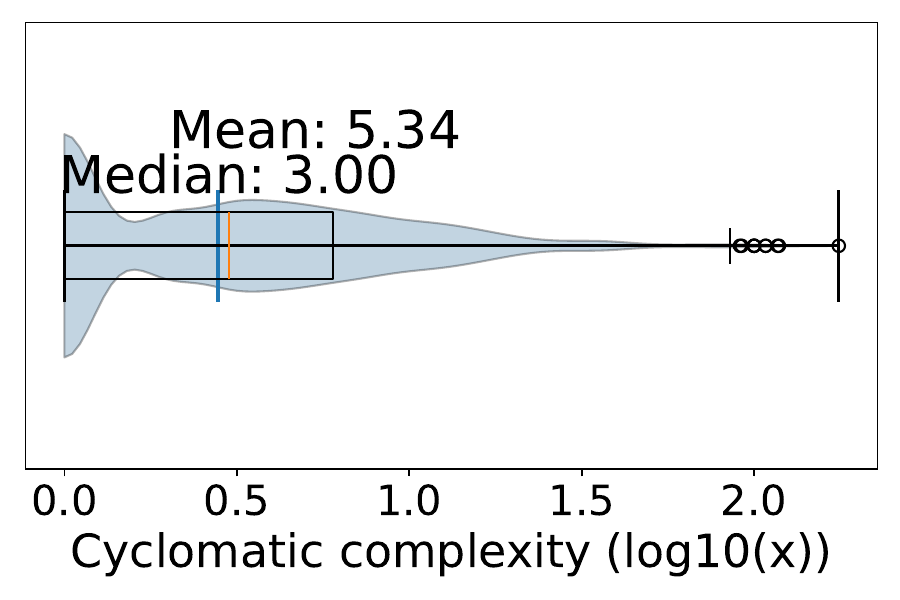}
  \end{minipage}
   \hfill 
  \begin{minipage}{.32\columnwidth}
    \includegraphics[width=\linewidth]{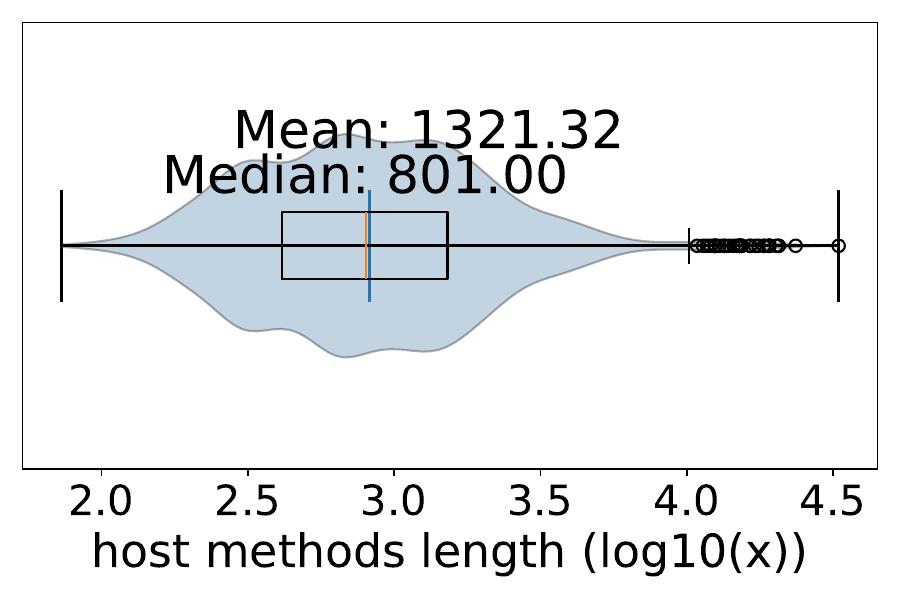}

  \end{minipage}
  \caption{Metrics for host methods in our corpus\label{fig:extended_corpus_metrics}}
\end{figure}

\subsection{Evaluation Metric: Recall@n with m\% tolerance\label{sec:metric}}\label{sec:datasets} 
Previous researchers~\cite{jextractsilva2015,gems2017ISSRE,REMS2023ICPC,Tsantalis2011} have designed metrics to evaluate \EM suggestion tools. In their evaluations, they analyze the top-n ranked suggestions generated by their tools at various $m\%$  tolerance levels. We also employ  \textit{Recall@$n$ with $m\%$ tolerance} to evaluate \tool.
First, to explain $m\%$ tolerance level, let us examine a specific instance.
Consider an extract method suggestion \(e_i(n_i, (s_i, e_i))\) where \(s_i\) and \(e_i\) are the start and end line numbers of the suggested code fragment to be extracted, while the oracle specifies \(e_j(n_j, (s_j, e_j))\) as the actual refactoring. 
The $m\%$ tolerance verifies whether the start and end line numbers of the suggestion are within the start and end lines of the ground truth extracted method, with a tolerance threshold at most $m\%$ of the length of the ground truth method, $L_i$. 
Formally, 
$|f_i - f_j| + |s_i - s_j| \le n/100 * L_i$.
Next, Recall@n is calculated as the percentage of host methods for which at least one of the top-n \EM suggestions matches the refactoring specified in the oracle, given $m\%$ tolerance level. 

{Given the non-deterministic nature of LLM outputs, recall rates for \tool may vary across invocations. However, the impact of this on \tool is minimal since we invoke the LLM multiple times. 
Despite the minimal effect, during the experiments, when we report the recall rate for \tool, we repeat the experiments $\mathcal{N}$ times and consider a distribution of recall rates, then report the \mean and standard deviation of these values. While a larger $\mathcal{N}$ is preferable, 
 we use $\mathcal{N}$ = 30, following the best practices suggested by Andrea et al.\cite{arcuri2014hitchhiker}.}

\end{enumerate}
\input{sections/04_RQ1}

\input{sections/04_RQ2}

\input{sections/04_RQ3}

\input{sections/04_RQ5}

%% file: sections/04_RQ1.tex
\subsection{\textbf{RQ1: Effectiveness of LLMs}\label{sec:rq1}}
Numerous LLMs have been developed by both open-source communities and proprietary companies~\cite{scao2022bloom,google_bard,gpt-4,anthropic-claude,llama-meta,falcon}.
Among them, 
\begin{enumerate*}[label=(\roman*)]
\item PaLM~\cite{google_bard},
\item GPT-3.5~\cite{gpt-3}, 
and
\item GPT-4~\cite{gpt-4}
\end{enumerate*}
are well-known and the largest LLMs, developed by Google and OpenAI, and we use them for the experiments.
While these models were not developed specifically for refactoring, they are versatile. We determine their effectiveness in generating \EM refactoring suggestions. 

\begin{figure}
\includegraphics[width=0.5\textwidth,keepaspectratio=true]{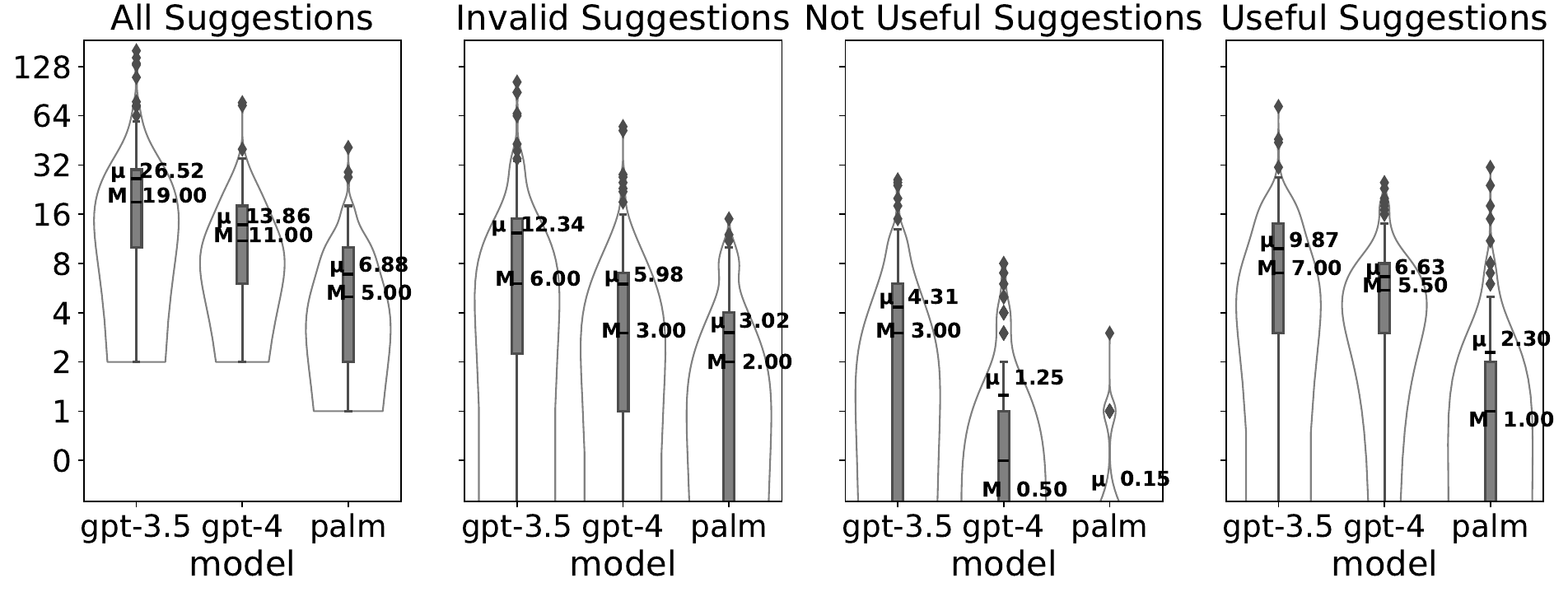}
     \vspace{-3mm}
    \caption{The capabilities of LLMs in generating refactoring suggestions. The plots show the number of suggestions per host method (notice the exponential scale)}
    \label{fig:rq1_llm_refac}
    \vspace{-3mm}
\end{figure}

\subsubsection{\textbf{Subject Systems and Experimental Setup}\label{sec:rq1_data}}
We employed a two-step process to assess the quality of refactoring suggestions generated by LLMs.

\textbf{Step 1:} 
We used \CommunityCorporaA and then prompted the LLM with each method to generate refactoring suggestions. 
To maximize the capabilities of LLMs to generate suggestions, we employed an iterative approach by adjusting the temperature parameter during LLM prompting, as explained in Section \ref{sec:gen_extensive}. 
Specifically, we conducted fixed-point iterations, repeatedly prompting the LLM until it no longer generated new suggestions for each temperature value from the set \{0, 0.2, 0.4, 0.6, 0.8, 1.0, 1.2\}. This iterative strategy allowed the LLM to create refactorings with varying levels of randomness and expand its search with each subsequent response to the prompt.
Then, we analyze the quality of the suggestions by quantitatively studying invalid suggestions (\Cref{def:invalid_extract_method}), and not useful suggestions (\Cref{def:notuseful_extract_method}).

\textbf{Step 2:} 
To significantly increase the validity of the observations made during Step 1, we employed the \ExtendedCorpus and prompted it with the best-performing LLM parameters from Step 1.
We then analyzed the quality of the suggestions, following the same process as in Step 1.

\subsubsection{\textbf{Results}}

In \Cref{fig:rq1_llm_refac}, box and violin plots show the distribution of suggestions per host method. Starting from the left, total suggestions, then those that are invalid, not useful, and finally, the useful suggestions generated by each LLM. 
The data reveals that all three studied LLMs excel in generating refactoring suggestions, with GPT-3.5, GPT-4, and PaLM averaging {27, 14, and 7} refactoring suggestions per host method, respectively. However, they also produced invalid suggestions with averages of 12, 6, and 3, respectively, and not useful suggestions with averages of 4, 1, and 0, respectively. Notably, GPT-3.5 yielded the highest number of useful suggestions, with an average of {10} per host method, compared to 7 and 2 averages for GPT-4 and PaLM, respectively. 
These results highlight the effectiveness of LLMs in generating refactoring suggestions while also highlighting the need for additional techniques to select only useful suggestions. 

To further bolster the validity of this claim, we selected GPT-3.5, the top-performing LLM in terms of applicable suggestions, and applied it to the \ExtendedCorpus.
Our observations, consistent with the findings in Step 1, revealed that GPT-3.5 generated total \llmNumberOfSuggestionsProducedOnExtendedCorpus refactoring suggestions, of which \errOrHallSuggestionsExtendedCorpus were hallucinations (i.e., invalid and non-useful), leaving only \applicableSuggestionsExtendedCorpus (\llmFinalUsefullPercentageExtendedCorpus) useful suggestions. 
This underscores the importance of employing post-processing techniques on LLMs' output, they cannot be used as-is for refactoring tasks.

\resultbox{
LLMs excel at generating \EM suggestions, yet only \llmFinalUsefullPercentageExtendedCorpus of these suggestions are useful.
}


%% file: sections/04_RQ2.tex
\subsection{\textbf{RQ2 : Sensitivity Analysis of \tool}\label{sec:rq2}}
We determine the optimal values for LLM hyper-parameters so that we can harness the full potential of LLMs when generating \EM refactorings. 
Then we study the contribution of our design decisions on enhancing the refactoring suggestions made by \tool.
This is crucial for revealing best practices for using LLMs for refactoring tasks, and it also enables us to quantitatively assess \tool's advancements over the raw performance of LLMs.

\subsubsection{\textbf{Subject Systems and Experimental Setup}\label{rq2_subjects}}
We use our oracle dataset, \CommunityCorporaA, to analyze the output of \tool by examining the effects of specific Temperature values ($t \in {0, 0.2, 0.4, 0.6, 0.8, 1.0, 1.2}$) and iteration numbers ($i \in {1, 2, \ldots, 10}$). For each parameter combination $(t, i)$, we invoke \tool to generate extract method suggestions for a host method within the dataset. In line with the methodologies adopted by previous researchers~\cite{jextractsilva2015,gems2017ISSRE,REMS2023ICPC,Tsantalis2011} for assessing earlier \EM automation tools, we evaluate each parameter setting by calculating Recall@5 with a 3\% tolerance level.
Here, to identify the optimal parameter values ($t$, $i$), we analyze a distribution of recall values (as explained in \Cref{sec:metric}) and select the mean recall.
Then, we assess the recall both with and without \tool's enhancements to the LLM suggestions. 

\begin{figure}[t]
        \centering
\includegraphics[height=3.25cm,keepaspectratio=true]{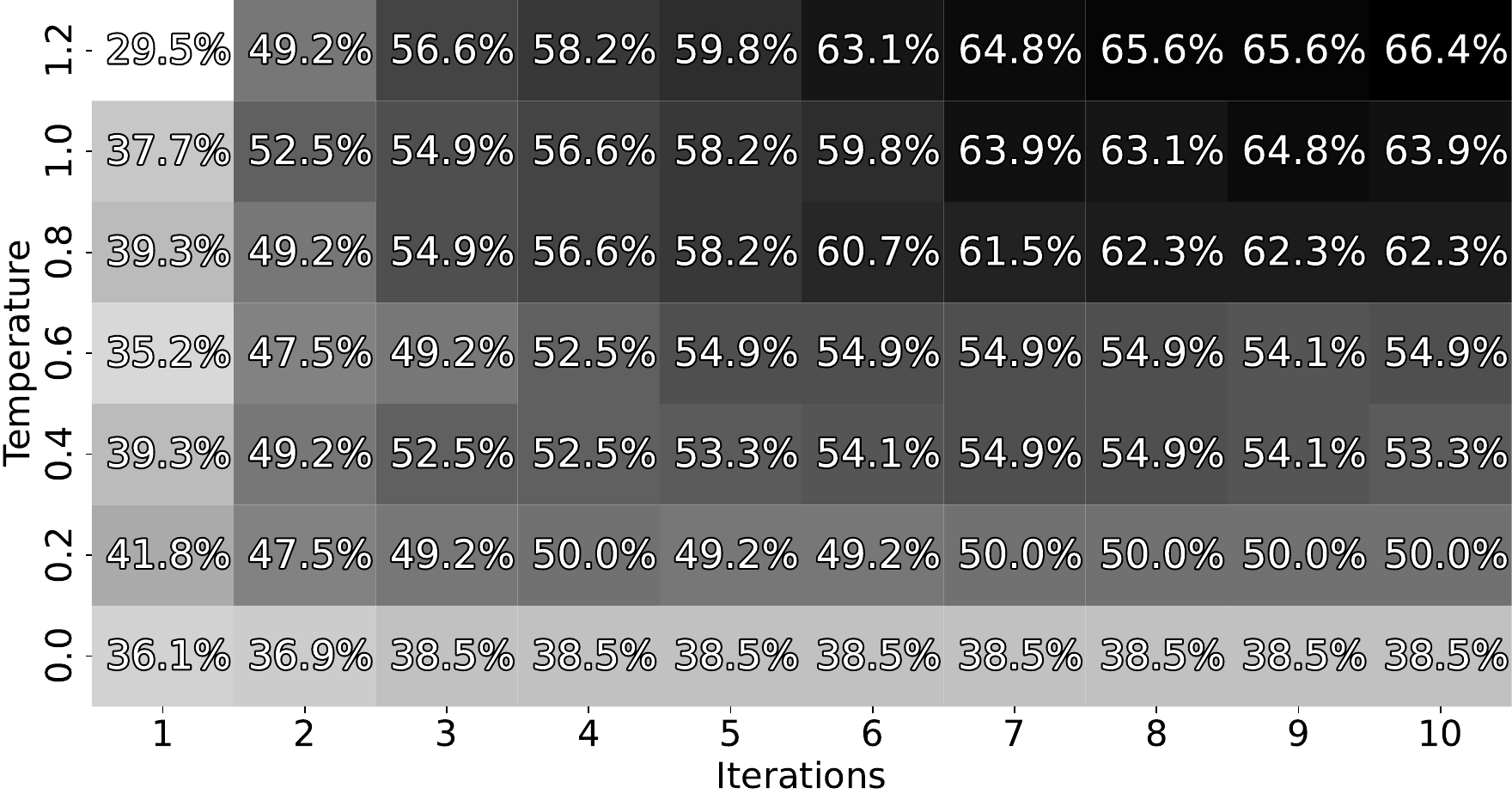}
    \caption{Change of Recall@5 along with Temperature and iterations for the \CommunityCorporaA}\label{fig:heatmap_corpusA} 
    \label{fig:empirical}
      \setlength{\tabcolsep}{-5pt} 

\end{figure}

\begin{table}
	\centering
	\caption{Ablation Study: Mean and Standard Deviation of Recall Values for different stages in \tool's pipeline.}
\fontsize{8pt}{9pt}\selectfont
	\label{table:ablation}
  \setlength{\tabcolsep}{3pt}
\begin{tabular}{l|ccc}
           & \multicolumn{3}{c}{Recall@5} \\
Method       & Tolerance         & Tolerance        & Tolerance      \\ 
       &   1\%         &  2\%        &  3\%       \\ 

\hline
Random rank, no heuristics       & $32.9\% \pm 2.3$          & $34.2\% \pm 2.3$         & $37.4\% \pm 2.4$        \\
Random rank, with heuristics       & $44\% \pm 2.8$         & $44.8\% \pm 2.9$        & $49.6\% \pm 2.9$       \\

Ranking and heuristics & \textbf{\emToolRecallOne} $\pm$ 1.6& \textbf{\emToolRecallTwo} $\pm$ 1.6& \textbf{\emToolRecallThree} $\pm$ 1.7    
\end{tabular}

\end{table}


\subsubsection{\textbf{Results}}
\Cref{fig:heatmap_corpusA} shows via heatmaps how \tool's mean recall varies with LLM  Temperature/Iterations.
The results show that higher temperature values and more iterations consistently yield superior results.
{Even though LLM randomness increases with higher temperature values, potentially leading to a greater variety of suggestions and possibly more hallucinations with more iterations, the enhancements implemented in \tool effectively refine and rank these suggestions in such a manner that its recall improves alongside these parameters.}
Notably, \tool recorded its peak Recall@5 scores at \emToolRecallThree, with the configuration set to a temperature of 1.2 and 10 iterations, establishing these parameters as the optimal settings for \tool's applications.

To further study the contributions of various components within the \tool's pipeline (see \Cref{fig:workflow}), we conducted an ablation study. 
We structured this into three distinct phases:
\begin{enumerate*}[label=(\roman*)]
\item In the initial phase, we processed the raw output from the LLM without any heuristic enhancements and randomly selected five candidates.
\item We considered the enhanced output with heuristics (see \Cref{sec:enhancing}) and arbitrarily choose five suggestions after enhancement.
\item We took into account the enhanced output and applied the tool's ranking mechanism.
\end{enumerate*}
These stages were instrumental in understanding the individual contributions of the various modules to the overall performance. 
As shown in \Cref{table:ablation}, the results emphasized that significant improvements were observed across all tolerance levels, with a notable enhancement at the 3\% tolerance level. 
Specifically, the baseline mean recall of the LLM was recorded at {37.4\%}. Through our design decisions, this figure was elevated to \emToolRecallThree. This enhancement vividly illustrates the substantial influence of the strategic decisions embedded in our tool's design.


\resultbox{
With higher temperature (1.2) and more iterations (10), LLMs produce better suggestions for \EMs. Our ranking and heuristics further amplified the quality of the suggestions, achieving an uplift of up to 26 percentage points.
}

%% file: sections/04_RQ3.tex
\subsection{\textbf{RQ3 : Effectiveness of \tool}}\label{sec:rq3}
\tool introduces a novel approach for refactoring recommendation that harnesses the power of LLMs. 
Thus, we assess its effectiveness relative to existing state-of-the-art tools.

\subsubsection{\textbf{Subject systems and Experimental Setup}}
We selected six tools that are representative of a wide array of techniques, including four (JDeodorant~\cite{Tsantalis2011}, JExtract~\cite{jextractsilva2015}, SEMI~\cite{Charalampidou2017}, and LiveRef~\cite{liferef2022ASE}) that use static analysis-based rules and software quality metrics, and two (REMS~\cite{REMS2023ICPC} and GEMS~\cite{gems2017ISSRE}) that use  prediction models based on machine learning techniques.

We first employ \CommunityCorporaA to evaluate the effectiveness of the selected tools. 
Following practices employed by the authors of other tools, 
we evaluate top-5 suggestions generated by the tools, calculating \mean Recall@5 at tolerance levels of 1\%, 2\%, and 3\% (for details on evaluation metrics refer to \Cref{sec:metric}). 
For \tool, we compute the metric for the best-performing hyper-parameters.
For four tools (GEMS, JDeodorant, JExtract, SEMI), we reuse the evaluation results reported by other researchers~\cite{gems2017ISSRE} because they used the same dataset and they defined the same metric formulation as we do.
For two tools (\textit{REMS} and \textit{LiveRef}), we had to run them ourselves: \textit{LiveRef} had not previously been executed on \CommunityCorporaA, and \textit{REMS} computes recall differently than any of the other tools. 
Notice that we contacted the authors and had extensive conversations with them to fully understand how to replicate their results best and how to use their tools so they perform the best.

Second, in order to bolster the robustness of our evaluation, we expand our evaluation to \ExtendedCorpus, which replicates \extendedCorpusSize \EM refactorings that took place in open-source projects. 
In this extended evaluation, we compare directly with {JExtract}.
During our evaluation on \CommunityCorporaA, we noticed that GEMS and JExtract had the highest recall rates following \tool, and surpassing all other tools. Their recall rates were remarkably similar, and given the extensive scale of the experiments and the time considerations, we opted for JExtract to execute and benchmark the recall rates against \tool, using a Recall@5 metric at 3\% tolerance.



\subsubsection{\textbf{Results}}
{\Cref{table:comparison} shows the comparative effectiveness of \tool in relation to the other six tools.
We tame the non-determinism of LLMs as detailed in \Cref{sec:metric}.}
Among the tools evaluated, GEMS ranked second highest, with JExtract closely behind. 
{\tool achieves superior performance with higher Recall@5 scores at 1\% and 3\% tolerance levels, recording values of \emToolRecallOne and \emToolRecallThree, respectively.}


To further strengthen the validity of our results, we applied both \tool and {JExtract} on the \ExtendedCorpus that includes \extendedCorpusSize actual refactorings from open-source projects.
For both tools, we calculated the recall for top-5 suggestions with a 3\% tolerance value.
As shown in \Cref{table:ExtendedComparison},
\textit{JExtract} achieved recall of \JExtractBestRecall, 
\textbf{while \tool's \mean recall was \ecBestToolRecall, which is \bestJetGPTOverBestJExtractRecall improvement over the baseline}.
{This shows that although \tool's advantage is modest when evaluated on a synthetic \CommunityCorporaA, its performance notably surpasses the baseline when it comes to replicating refactorings actually performed by developers.}

{To address the non-determinism of LLM outputs, we statistically analyzed the recall values of \tool and JExtract over the \ExtendedCorpus. 
We created a distribution of recall values for \tool and then employed the Wilcoxon Signed-Rank Test to compare this distribution against the recall values of JExtract. 
The test rejected the null hypothesis, indicating a statistically significant difference in recall. 
Following this, we used the Hodges-Lehman estimator on each combination of distributions to quantify the difference in recall rates. The resultant value was 12.4 percentage points, suggesting a notable difference in performance between the two tools. This shows that \tool's improvements over the baseline are statistically significant. }


\begin{table}[t]
	\centering
	\caption{Evaluation results of \tool with respect to six other tools on \CommunityCorporaA}
\fontsize{8pt}{8pt}\selectfont
	\label{table:comparison}
 
\begin{tabular}{l|ccc}
           & \multicolumn{3}{c}{Recall@5} \\
Tool       & Tolerance 1\%         & Tolerance 2\%        & Tolerance 3\%       \\ \hline
REMS       & 1.6\%          & 1.6\%         & 1.6\%        \\
GEMS       & 54.2\%         & \textbf{59.8\%}        & 62.6\%       \\
JDeodorant & 14.8\%         & 18.4\%        & 23.8\%       \\
JExtract   & 52.2\%         & 59.3\%        & 61.9\%       \\
SEMI       & 38.0\%         & 47.0\%        & 55.5\%       \\
LiveREF    & 10.6\%         & 10.6\%        & 13.1\%       \\
\tool & \textbf{\emToolRecallOne} $\pm$ 1.6 & \emToolRecallTwo $\pm$ 1.6 & \textbf{\emToolRecallThree} $\pm$ 1.7

\end{tabular}

\end{table}

\begin{table}
	\centering
	\caption{Evaluation results of EM-Assist with respect to six
other tools on \ExtendedCorpus of \extendedCorpusSize instances}
\fontsize{9pt}{9pt}\selectfont
	\label{table:ExtendedComparison}
  \setlength{\tabcolsep}{3pt}
\begin{tabular}{l|ccc}
           & \multicolumn{3}{c}{Recall@5} \\
Tool       & Tolerance 1\%        & Tolerance 2\%         & Tolerance 3\%     \\ 
\hline
EM-Assist      & $52.2\% \pm 0.9$          & $52.7\% \pm 1$          & \ecBestToolRecall $\pm 1$      \\
J-Extract      & 38.8\%         & 39.3\%        & \JExtractBestRecall      \\
   
\end{tabular}

\end{table}

\resultbox{
When using an oracle of synthetic refactorings, \tool has a slightly superior recall rate compared to its peers. When using an oracle of real-world refactorings, \tool's recall rate improvements over its peers are  \bestJetGPTOverBestJExtractRecall, showing it better aligns with developer preferences. }


%% file: sections/04_RQ5.tex
\subsection{\textbf{RQ4: Usefulness of \tool}\label{sec:rq5}}

\tool is an interactive tool that provides maximum automation while still taking into account human input (see~\cite{tool:Marketplace}).
It presents refactoring suggestions to a developer, and based on their selection, \tool applies the chosen refactorings and changes the code. 
Therefore, it is important to study the usefulness of these suggestions to developers and to gain insight into 
their decision-making process. 
We will take into account this feedback when releasing future versions. 

\subsubsection{\textbf{Data and Experimental setup}}
To study the usefulness of the refactoring suggestions, we use the Firehouse survey research method~\cite{firehouse2015TSE}.
A phenomenon and its solution are studied right after it happens (\textit{e.g.}, observing victims' behavior right after a house catches fire). 
Accordingly, we engage with open-source developers immediately after they commit a long method into their repository, and we present them with suggestions for \EM generated by \tool. 
This direct interaction with developers, leveraging their recent familiarity with the code, ensures they provide reliable answers and are best equipped to evaluate the quality of suggestions.
Notably, this method, as demonstrated by Silva et al.~\cite{WhyWeRefactorFSE2016}, boasts a significantly higher response rate compared to other survey-based studies that require developer participation.

To perform firehouse surveys, we engaged with developers contributing to the open-source projects \textit{JetBrains/IntelliJ-Community} Edition~\cite{IntelliJCE} and \textit{JetBrains/Runtime}~\cite{JetBrainsRuntime}.
The IntelliJ Community Edition, written in Java and Kotlin, has 15.8k stars, 417k commits, and 952 contributors. JetBrains/Runtime, written in Java, has 969 stars, 76k commits, and 806 contributors. These projects exemplify significant engagement, maturity, and attention to code quality, making them ideal for our study.


\textbf{Firehouse Survey Method:}
We monitored the projects daily, analyzing every new commit to identify newly added long methods. To enhance this, we extended RefactoringMiner~\cite{refactoringminer2} to prevent mistakenly identifying refactored methods (\textit{e.g.}, renamed or moved) as newly added.
Once we identified a long method, we contacted the developer.  
To minimize the intrusion for the developers, 
we only asked their opinion for one single long method (regardless of how many methods they authored) and only presented three suggestions for their method (regardless of how many \tool generated). 


Then we asked developers for their level of agreement with each suggestion on a 6-point Likert scale ranging from Strongly Agree to Strongly Disagree.
Additionally, we encouraged them to share their reasoning, detailing what aspects they liked or disliked about the suggestions or any changes they would consider.
{As developers answered with free-form text, we conducted thematic analysis to interpret these responses, adhering to established best practices~\cite{CRUZES2011440,Virginia2006,campbell2013coding}.}


\subsubsection{\textbf{Results}}

We surveyed \totalSent developers, out of which \totalResponse responded, bringing us to a \responserate response rate. 
This response rate is notably higher than the typical response rate in questionnaire-based software engineering surveys, which typically hovers around 5\%~\cite{Singer2008}, and this can be attributed to using firehouse surveys.
\Cref{tab:agreement_levels} shows the agreement levels. For each developer survey, we present the entry for the highest-rated suggestion by the developer. 
\Cref{tab:agreement_levels} shows that even when giving just three ranked suggestions per method so that we do not overwhelm the developer, \agreeratio of respondents find them useful (i.e., chose a positive rating).
Thus, \tool generates useful refactorings that align with developer preferences. 
As \tool finds improvement opportunities and generates useful suggestions even for high-quality projects like the ones we used, we are confident it would discover useful refactoring opportunities in regular-quality projects.



\begin{table}[h]
    \centering
    \caption{Developers' levels of agreement to the refactoring suggestions produced by \tool \label{tab:agreement_levels}}
    \resizebox{\columnwidth}{!}{
    \fontsize{9pt}{12pt}\selectfont 
    \begin{tabular}{cccccc}
        Strongly  & Disagree & Somewhat  & Somewhat  & Agree & Strongly  \\
         Disagree &  &  Disagree &Agree   &  & Agree  \\
           \hline
        \SD & \D & \SWD & \SWA & \AG & \SA \\
    \end{tabular}
    }
\end{table}

Developers remarked that these refactoring suggestions significantly enhance code quality.
They appreciated a fresh perspective on their code: ``\textit{…these suggestions made me look at this code with new eyes once more, and I will try to refactor it}''.
Furthermore, developers strongly desire to see \tool in production:  ``\textit{Thank you for interesting suggestions! Hope to see this in production in the future.}''
These encouraging comments highlight the positive impact and usefulness of \tool's recommendations in the daily development process.



\noindent The top-3 reasons why developers did not accept suggestions: 
\begin{enumerate}[wide, leftmargin=*,labelwidth=!, labelindent=0pt]
    \item  Splitting a monolithic algorithm into smaller parts could potentially complicate code organization, and the original method is concise enough. It suggests that in the future, \tool should more accurately take the developer's method size preferences into consideration.
    \item The extracted method has too many parameters. This shows the importance of implementing additional filters based on the number of parameters in the suggestion. 
    \item The extracted method does not promote reuse: While this is a valid concern, it is important to note that \tool focuses solely on individual methods as input and does not take into account broader contextual factors, such as a file or project-level considerations. In the future, we could extend \tool's scope to suggest extract methods for reuse at higher levels, such as file, module, or project levels~\cite{alomar2023anticopypaster-approach, alomar2023anticopypaster-tool}.
\end{enumerate}


\resultbox{
Developers say 81\% of \tool's suggestions are useful. 
}

%% file: sections/06-threats-to-validity.tex
\section{Threats To Validity}
\begin{enumerate}[wide, leftmargin=*,labelwidth=!, labelindent=0pt]
\item \textbf{Internal Validity}: \textit{Does our work produce valid results?}
\tool's effectiveness can vary depending on the oracle used for evaluation. 
To mitigate this, we employ two datasets for assessment: a publicly available, independent dataset commonly utilized by other researchers, and a more extensive dataset derived from open-source repositories. Our thorough evaluation of \tool on these datasets reveals its enhanced recall capabilities, outperforming other state-of-the-art tools.
Furthermore, the firehouse survey results also show that open-source developers found \tool's suggestions useful.

\item \textbf{External Validity}: \textit{Do our results generalize?} 
\toolit's effectiveness hinges on the chosen LLM model.
As LLMs continue to advance, especially with upcoming iterations trained on more extensive datasets, we anticipate further improvements in outcomes.
The architecture of \tool allows for effortless adaptation to future LLMs.
Lastly, it is important to highlight that while our approach is conceptually language-agnostic, our current \tool implementation is tailored for Java and Kotlin. Therefore, we cannot extrapolate our performance results to programs written in other languages.


\item \textbf{Verifiability}:
We make \tool, the datasets and results, publicly available~\cite{replicationPackage}. 
\end{enumerate}

%% file: sections/05-relatedWork.tex
\section{Related Work}\label{sec:related}
We observed two primary approaches for \EM refactoring: (i) tools designed to recommend specific code fragments for developers to refactor using \EM, and (ii) tools that predict whether a host method needs \EM refactoring. Accordingly, the related work is structured into two sections.


\noindent\textbf{Suggesting EM refactoring:}
Several studies~\cite{negara2013,murphy2012,refactoringminer2,murphy2008breaking} indicate that EM refactoring is among the top five most frequently performed practices, leading to the development of numerous supporting tools.
JDeodorant~\cite{Tsantalis2011} automatically calculates block-based program slicing for variables in assignment statements. Given a slicing criterion as input, JDeodorant can extract relevant statements while preserving program behavior. JExtract~\cite{jextractsilva2015} operates based on the block structure, where each block structure contains a group of statements organized with a linear control flow. Given a method, JExtract detects involved block structures and heuristically ranks them as refactoring candidates. SEMI~\cite{Charalampidou2017} explores the coherence between statements and returns code fragments with high cohesion as refactoring candidates. LiveRef~\cite{liferef2022ASE} is an IntelliJ plugin that offers real-time \EM refactoring suggestions and immediate application, guided by code quality metrics and visual cues in the code editor. It continually adapts to code changes, providing live refactoring support.
REMS~\cite{REMS2023ICPC} utilizes multi-view representations from the code property graph. It then trains a ML classifier to guide the extraction of suitable lines of code as a new method.
GEMS~\cite{gems2017ISSRE} encodes metrics related to complexity, cohesion, and coupling as features to train ML classifiers for recommending EM refactoring opportunities.
Other tools propose EM refactoring based on program slicing~\cite{lakhotia1998restructuring, Maruyama2001, abadi2009fine}, separation of concerns, and the single responsibility principle~\cite{Charalampidou2017}, while some~\cite{tairas2012increasing, yue2018automatic, yoshida2019proactive, alomar2023anticopypaster-approach} aim to reduce code clones and duplication.

In contrast, our work pioneers the integration of LLMs with static analysis techniques to scrutinize and refine the outputs of LLMs, compensating for their lack of understanding in program semantics, thereby facilitating the correct and safe execution of EM refactoring.

\noindent\textbf{Predicting EM Refactoring:} 
Researchers identified the necessity for implementing EM refactoring.
Aniche et al.~\cite{aniche2020effectiveness} evaluated the effectiveness of various ML algorithms for predicting software refactorings, including EM refactoring. 
Their models were trained on thousands of refactorings mined with RefactoringMiner~\cite{refactoringminer2} from open-source projects. Van der Leij et al.~\cite{van2021data} replicated the study~\cite{aniche2020effectiveness} at the ING company and discovered that ML models predict very well the opportunity for applying EM refactoring. 
Others~\cite{alomar2022documentation,sagar2021comparing} use commit messages to predict software refactorings. 
While these tools are very good at predicting the type of refactoring a method needs to undergo, they complement nicely with \tool and its ability to suggest code fragments to be extracted and also to apply the refactoring.

%% file: sections/06-conclusion.tex
\section{Conclusions}\label{sec:conclusions}
\tool is the first refactoring tool that exploits the untapped potential of LLMs for refactoring tasks. 
AI systems can break down at unexpected places. 
For an LLM, this results in refactoring suggestions that seem plausible at first reading but are actually deeply flawed.   Our experiments show that LLMs are not reliable and need to be checked. We discovered a novel way of checking LLM results and making them useful for refactoring tasks.


\tool recommends \EM refactorings that align with developers' preferences. This is evidenced when replicating thousands of real-world refactoring scenarios from open-source repositories. Moreover, our firehouse surveys with developers of high-quality codebases that authored recent changes revealed that \agreeratio of respondents found \tool's suggestions useful.

We discovered a new set of best practices when using LLMs for refactoring. 
One of the biggest challenges with LLMs is taming non-determinism. Unlike in traditional static analysis tools where we avoid redundant computations, 
working effectively with LLMs requires re-prompting  (asking the same question several times) and creating a superset of all suggestions. This was key to get the most out of the LLM. Thus, we designed novel ranking methods to take into account the LLM workflow. Moreover, we learned that the more precise the prompt, the higher the quality of the suggestions. Few-shot learning worked best for refactoring tasks. 


\tool provides maximum automation of the full lifecycle of employing an LLM (i.e., prompting, validating, and enhancing suggestions) and it executes refactorings correctly within the IDE while still keeping the programmer in the loop as the ultimate decision maker. This ushers a new era when AI assistants do not take over the programmers, but become effective companions for code renovation tasks. Together, programmers, assisted by AI and the IDE, go further. 

In this work, we focused on \EM refactoring for Java and Kotlin, but our approach can be expanded to other refactoring automation tasks, and to other programming languages.
We hope that the best practices that we discovered for using LLMs effectively for refactoring tasks inspire others to further advance the field of refactoring. 

\section{Open Science Policy}
We have made \tool and evaluation data available anonymously on our website~\cite{replicationPackage}, and the tool is freely available to be installed from the JetBrains Marketplace~\cite{tool:Marketplace}.

\section{Acknowledgements}
{We thank the ML Methods in Software Engineering Lab at JetBrains Research, and the FSE-2024 reviewers for their insightful and constructive feedback for improving the paper.
This research was partially funded through the NSF grants CNS-1941898, CNS-2213763, and the Industry-University Cooperative Research Center on Pervasive Personalized Intelligence.}